\documentclass[runningheads]{llncs}
\usepackage[T1]{fontenc}
\usepackage{algorithm}

\usepackage{algorithmicx}
\usepackage{todonotes}
\usepackage{amsmath}
\usepackage{svg}
\usepackage[most]{tcolorbox}
\usepackage[font=small,skip=2pt]{caption}
\usepackage{float} 
\usepackage{balance}
\usepackage{graphicx}
\begin{document}

\newtcolorbox{boxA}{
    fontupper = \scriptsize\bf,
    boxrule = 1pt,
    colframe = black 
}
\title{An AUTOSAR-Aligned Architectural Study of Vulnerabilities in Automotive SoC Software}
\titlerunning{SoC Software Security: Deep Dive}
%
\author{Srijita Basu\inst{1}\orcidID{0000-0002-6835-947X} \and
Haraldsson Bengt\inst{2}\orcidID{0009-0005-3219-8677} \and
Miroslaw Staron\inst{1}\orcidID{0000-0002-9052-0864}  \and
Christian Berger\inst{1}\orcidID{0000-0003-3929-0925} \and
Jennifer Horkoff\inst{1}\orcidID{0000-0002-2019-5277}  \and
Magnus Almgren\inst{1}\orcidID{0000-0002-3383-9617}}
\authorrunning{S. Basu et al.}
%
\institute{Chalmers University of Technology
and University of Gothenburg,
417 56 Göteborg, Sweden
\email{\{srijita.basu,miroslaw.staron,christian.berger,jennifer.horkoff\}@gu.se, magnus.almgren@chalmers.se}\\
\and
Scania CV AB, Södertälje, Sweden\\
\email{bengt.haraldsson@scania.com}
}
\maketitle              
\begin{abstract}

Cooperative, Connected and Automated mobility (CCAM) are complex cyber–physical systems (CPS) that integrate computation, communication, and control in safety-critical environments. At their core, System-on-Chip (SoC) platforms consolidate processing units, communication interfaces, AI accelerators, and security modules into a single chip. AUTOSAR (AUTomotive Open System ARchitecture) standard was developed in the automotive domain to better manage this complexity, which defines layered software structures and interfaces, and to facilitate reuse of HW/SW components. However, in practice, this integrated SoC software architecture still poses security challenges, particularly in real-time, safety-critical environments. Recent reports highlight a surge in SoC-related vulnerabilities, yet systematic analysis of their root causes and impact within AUTOSAR-aligned architectures is lacking. This study fills that gap by analyzing 180 publicly reported automotive SoC vulnerabilities, mapped to a representative SoC software architecture model that is aligned with AUTOSAR principles for layered abstraction and service orientation. We identify 16 root causes and 56 affected software modules, and examine mitigation delays across Common Weakness Enumeration (CWE) categories and architectural layers. We uncover dominant vulnerability patterns and critical modules with prolonged patch delays and provide actionable insights for securing automotive CPS platforms including guides for improved detection, prioritization, and localization strategies for SoC software architectures. strategies in SoC-based vehicle platforms. 



\keywords{Architecture \and Automotive  \and SoC \and Vulnerability .}
\end{abstract}
\section{Introduction}

 
Secure software architecture plays a critical role in modern Cooperative, Connected and Automated mobility (CCAM), where increasing functionality, ranging from performance optimization and infotainment to autonomous driving, demands the continuous evolution of automotive software systems. Automotive System-on-Chips (SoCs) have always been integral to the CCAM ecosystem, integrating multiple functions into one chip. Historically, micro-controllers handled isolated tasks such as engine control and braking. Over the past two decades, SoCs from vendors like NVIDIA, Qualcomm, and Intel have transformed from providing normal infotainment and safety features to high-performance AI-driven processors, enabling autonomous driving, advanced infotainment, and other connectivity features. This shift has brought significant architectural complexity. The number of software modules, their layered interactions, and the growing functional integration within SoCs pose new challenges for system-level analysis and security.


Because of the open ecosystem where both automotive-specific and general modules are used in the automotive software, the security of automotive CPS has become a pressing concern. Recent incidents involving data leakage, remote control exploits, and service disruptions \cite{ref_1} have highlighted pervasive risks. Prior work  \cite{ref_20} shows that nearly 67\% of automotive software vulnerabilities reported in the past eight years relate directly to SoC-based systems. Given the integration density of modern SoCs, a vulnerability in a single function can potentially compromise the entire stack.



While AUTOSAR provides a standardized reference architecture, its specifications, especially in the classic and adaptive platforms, are abstract, platform-agnostic, and not designed for direct mapping of real-world vulnerabilities. In contrast, vulnerabilities are typically reported at the code level, tied to specific drivers, firmware, or system service software. Therefore, we first derived an intermediate architecture model by examining open-source SoC software repositories to understand how real modules (e.g., WLAN drivers, IPC services, HALs) are structured and deployed. We then aligned this architecture with AUTOSAR concepts to retain compatibility. We mapped a curated subset of publicly reported automotive SoC software (ASoCS) vulnerabilities from the National Vulnerability Database (NVD) to this representative architecture model. Despite the limited availability of open automotive repositories, our study reveals critical trends and patterns through the following set of focused research questions (RQs). 

\begin{itemize}
    \item \textbf{RQ1}: What are the root causes behind the ASoCS vulnerabilities, and to what extent can these root causes explain the identified vulnerabilities?
    
    \item \textbf{RQ2}: How are the vulnerabilities distributed across different architectural components/modules of the ASoCS architecture model stack?
    
    \item \textbf{RQ3}: How does the vulnerability mitigation time vary across different CWEs and ASoCS modules?
\end{itemize}

We make the following contributions in the context of ASoCS architecture and its vulnerabilities -- 1) We use a literature-informed ASoCS architecture model aligned to AUTOSAR to study the exposure of architecture layers to real-world vulnerabilities, 2) We present different findings on ASoCS vulnerabilities by answering the above RQs and 3) We discuss CPS and AUTOSAR specific use cases to reflect on our findings. In particular, our findings allow software architects and security test leaders to identify risk-prone modules faster, prioritize secure design efforts, and improve vulnerability analysis. Additionally, our component-level vulnerability mapping can aid in making more informed decisions during component selection and procurement, contributing to overall software quality and security posture in SoC-based vehicle platforms.

\section{Related Work}
\label{sec:related_work}
We highlight research that analyzes automotive CPS software and their security. Some of them are analytical studies while some propose real solutions. 

An exploratory study by Garcia et al. \cite{ref_3} analyzed automotive software bugs for two of the most popular open-source software projects for autonomous driving: Autoware and Baidu Apollo. They studied 499 bugs and reported the root causes for the same and also identified the affected software components. 


Mashkoor et al. \cite{ref_4} presented a systematic mapping of model-driven engineering for safety and security software systems. 29 research articles (highest) in the automotive domain were selected by the authors. 11 of these were found to be a part of the architecture and design phase. Upon inspecting these papers, we could get a very few ones using reference architecture in their studies. One of these papers used the Evita automotive reference architecture for the in-vehicle network. No papers with automotive SoC software architecture mapping was found.

Basu and Staron  \cite{ref_20} presented an empirical study on automotive software vulnerability for the span 2018-2024, mainly focusing on the distribution of CVSS score, attack vectors and CWE categories over the 1663 identified vulnerabilities.

Bella et al.~\cite{ref_5} propose CINNAMON, a Basic Software Module (BSW) built on the AUTOSAR Classic Platform that enhances in-vehicle communication security by adding confidentiality to the existing integrity and authentication guarantees of the standard Secure Onboard Communication (SecOC) module. Unlike SecOC, which does not provide encryption, CINNAMON integrates cryptographic protection, freshness validation, and configurable security profiles. The authors implement and evaluate the module on STM32-based ECU prototypes with minimal performance overhead. Their work complements our vulnerability-centric analysis by showcasing how secure module design can be embedded within AUTOSAR's communication stack.


While prior research provides valuable insights into automotive software security, such as bug classification, CVE analysis, or model-driven safety techniques, most studies focus either on individual repositories or offer high-level analyses without architectural grounding. Our mapping approach aligns vulnerabilities to functional blocks within an AUTOSAR-aligned architecture, offering practical, actionable insights grounded in real-world software structures. By organizing and analyzing vulnerabilities across well-defined layers and modules within a representative SoC software stack, our study facilitates targeted vulnerability localization and deeper root cause identification. This architectural perspective supports industry practitioners such as security testers, software architects, and procurement teams in pinpointing risk-prone modules, anticipating mitigation delays, and making informed decisions around component selection and software quality assurance.

\section{Methodology: Architecture Modeling, Vulnerability Data Collection and Mapping}
\label{sec:methodology}

The method followed for this study can be divided into three parts. First, the ASoCS architecture model was derived as depicted in \emph{Step 1}. Next, the ASoCS vulnerabilities i.e. CVEs (Common Vulnerabilities and Exposures) \cite{ref_20} were collected, and the ones with open-source code were identified in \emph{Step 2}. Finally, in \emph{Step 3} the method for vulnerability mapping with i) root causes, ii) architectural modules and iii) mitigation time is presented.


Any manual analysis or decisions involved a group of two researchers, one from industry and another from academia. We used Cohen's Kappa coefficient for agreement between the two researchers and achieved a score of 0.81, indicating a high level of consistency in our manual process. 


\subsection{Step 1: Automotive SoC Software Architecture Model}
\label{subsec:step1} 

 {An inductive, literature-informed approach, based on widely used automotive and embedded platforms and standards, was used to derive a representative Automotive SoC Software (ASoCS) architecture model for this analysis. The model was constructed through a detailed investigation of open-source repositories including the Android Open Source Project (AOSP), Code Aurora Forum (CAF), and Qualcomm Board Support Package (BSP) layers \cite{ref_6} \cite{ref_7}. These repositories were studied to understand the functional decomposition and file organization of SoC components. Directory structures, dependency graphs, and interface definitions were analyzed to extract recurring architectural patterns and software boundaries. To ensure structural fidelity with industry practices, we aligned the extracted architectural layers with AUTOSAR, which defines a service-oriented and layered architecture model for modern automotive platforms. Fig.~\ref{fige1} (Full Software Architecture) and Fig.~\ref{fige2} (Kernel and User Space of Central Processor Module) present a compact view of our architecture model. The key derivation principles were as follows:

\begin{itemize}
    \item \emph{Layered Decomposition of the Central Processor (CP)}: The CP is split into User Space, Kernel Space, and Secure Execution Environment (SEE) to mirror Android/Linux system layers found in embedded automotive environments. Moreover, the separation between user and kernel space in our model aligns with AUTOSAR's layered abstraction \cite{ref_8}, where application-level functionalities (adaptive applications) are decoupled from low-level execution services such as the POSIX-based operating system and execution management modules provided within the foundation layer.

     \item \emph{Function based Subsystem Isolation}: Digital Signal, Communication, Graphics Processing, etc. modules are separated based on platforms like Snapdragon Automotive and NVIDIA DRIVEs. Each has firmware, drivers, and function-specific APIs interacting with the CP but running semi-independently. Several of these modules like the Communication, Input-Output Memory Management Unit (IOMMU), Inter-Process Communication (IPC), etc.~correspond to the basic software modules defined in the AUTOSAR Platform \cite{ref_9}. However, to capture real-world SoC integration patterns, our model also includes additional modules such as Neural Processor and Graphics Processor, which are not explicitly specified or abstracted in AUTOSAR. This deviation is justified by the increasing integration of AI accelerators and advanced GPU-based infotainment systems in automotive SoCs. This hybrid approach allows us to maintain general alignment with AUTOSAR's service-oriented principles while accommodating the architectural diversity found in commercial SoC platforms.
     
    \item \emph{Security Domain Recognition}: Components such as Secure Execution Environments (SEE) are retained explicitly in our model to reflect Trusted Execution Environment (TEE) implementations common in Qualcomm and ARM-based SoCs. While AUTOSAR mentions security services, it typically treats TEEs as part of the hardware abstraction or middleware, without modular representation. We chose to include these as distinct modules due to their direct relevance in vulnerability localization and threat modeling.

    \item \emph{Abstract Vendor-specific Labels}: Special care was taken to abstract vendor-specific labels (e.g., ``ADAS'') while preserving functional categories such as audio processing, inter-processor communication, and power management etc., services. 
    
\end{itemize}

Though the functionalities of some of these modules and sub-modules are discussed while the results are presented, the detailed description of each is outside the scope of this paper. Table~ \ref{tab0} presents an alignment of the ASoC architecture with AUTOSAR. 
\vspace{-0.3cm}
\begin{table}[htbp]
\caption{Functional Alignment of Derived SoC Architecture with AUTOSAR Concepts} 
\tiny
\begin{center}
\begin{tabular}{|p{0.20\linewidth} | p{0.30\linewidth} | p{0.40\linewidth} |}
\hline
\textbf{AUTOSAR Concept}  & \textbf{Mapped ASoCS Architecture Component} &  \textbf{Justification}\\
\hline
Application Layer  & Applications in User Space (Camera, Audio, AI Apps) & These represent functional services exposed to the vehicle, such as driver assistance, media playback, and AI tasks. \\
\hline
Runtime Environment (RTE) & Frameworks \& HALs (e.g., Wi-Fi HAL, Audio HAL) & HALs abstract hardware-specific APIs and enable portable application logic, similar to the service mediation role of AUTOSAR's RTE. \\
\hline

Basic Software (BSW) & Kernel-space services and complex device drivers (WLAN, GPU, IPC, etc.) & Implement platform services, resource management, and execution control, similar to BSW in AUTOSAR\\
\hline
Microcontroller Abstraction Layer (MCAL)  & Hardware-specific drivers and firmware like DSP, Video processor drivers, NPU etc. & Interface with hardware accelerators providing lowest-level hardware access much like how MCAL provides hardware abstraction in AUTOSAR Classic.\\
\hline
Security Extensions  & Secure Execution Environment (SEE) and Secure Processing Module (SPM) & These ensure cryptographic services, and secure enclave operations—aligned with AUTOSAR's security architecture and policy-based access (SecOC, IAM, etc. \cite{ref_92}).\\






\hline
\end{tabular}
\label{tab0}
\end{center}
\end{table}
\vspace{-0.6cm}


\begin{figure}[ht]
\begin{center}
\includegraphics[scale=.25]{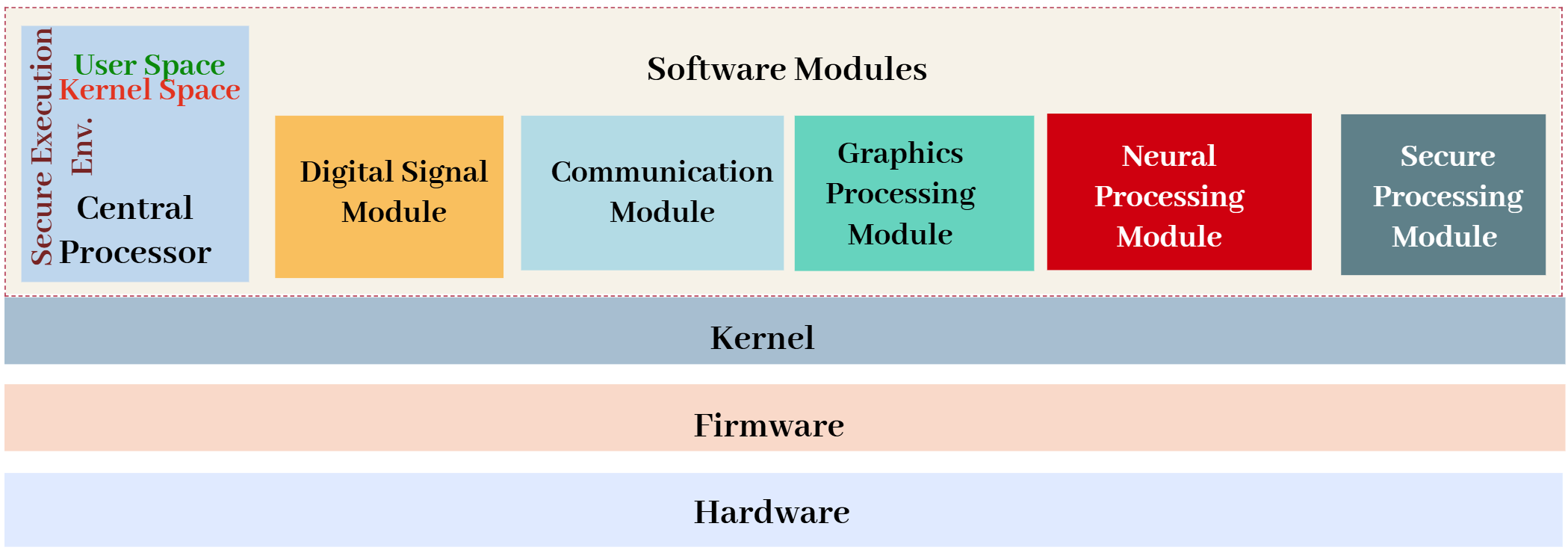}
\caption{Automotive SoC Software Architecture Model }
\label{fige1}
\end{center}
\end{figure}

\begin{figure}[ht]
\begin{center}
\includegraphics[scale=.25]{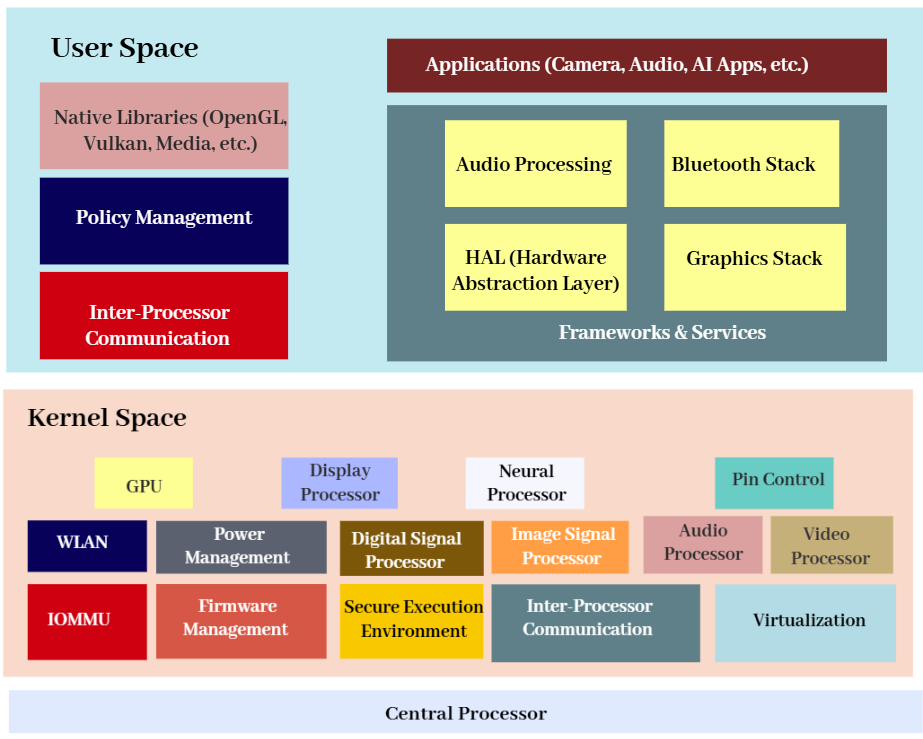}
\caption{ASoCS, Central Processor: Kernel and User Space}
\label{fige2}
\end{center}
\end{figure}
As we can see in Fig.~\ref{fige2} the CP runs the general purpose code and controls sending of tasks and managing communication between itself and the other specialized processors (Digital Signal Module (DSM), Graphics Processing Module (GPM), etc.). For instance, if we consider the use case scenario of \emph{detecting the slowing down of a nearby vehicle and alerting the driver} using the model above, then the steps would include, i) The camera sensors capture a video feed and send the data to the \emph{Image Signal Processor (ISP)} sub-module of the \emph{CP (kernel space)}, ii) After checking various factors such as: data is not empty, all pixels are fully saturated, etc., the ISP driver sends the data to the \emph{DSM module} for real-time object detection, iii) The \emph{DSM} uses its AI models for real-time object detection (tries to determine whether the other vehicle has actually slowed down), iv) As soon as the vehicle slow-down is detected by the \emph{DSM}, it communicates the same to the \emph{CP Kernel driver for GPM}, which in turn sends the message to the \emph{GPM, firmware} and v) The \emph{GPM} uses its drivers to render a warning on the infotainment 

\subsection{Step 2: Automotive SoC Software CVE Collection and Filtration}
\label{subsec:step2}

The process involved the following steps:

\begin{enumerate}

\item \emph{Download CVE List}: A year-wise CVE list (2018 to 2025, February) was downloaded from the NVD in JSON format.
and imported into a database. Details of the step can be found in our previous work \cite{ref_20}.


\item \emph{Filtering ASoCS CVEs - Phase 1}: We executed a search query based on \emph{CVE description text} and \emph{Common Platform Enumeration (CPE)} to filter the ASoCS vulnerabilities. The Keywords, \textbf{Auto, SoC, Qualcomm, Automotive, Texas, NXP, Renesas, NVIDIA, Snapdragon} were used to filter on the \emph{CVE description text} and chipset names like \textbf{QCA6574AU, QCA6595AU, QCA6584AU, etc.)} were used to filter on \emph{CPE}. The complete list can be found in\footnote{\url{https://github.com/SriAbir/AutomotiveSoCSoftware}}.

This dual search-based query was beneficial as sometimes the CVE description might not contain direct keywords indicating its association with ASoCS. In all such cases, the chipset name from the CPE description gave a relevant result. Again, there can be instances where the CPE did not register all valid chipsets that were affected, and here, the CVE description can compensate. The details of writing an optimized query in this step can be found in our previous work \cite{ref_20}. \textbf{1,113} vulnerabilities were filtered at the end of this step.

\item \emph{Filtering ASoCS CVEs - Phase 2}: After getting this list of vulnerabilities, 
we further examined them to check for the availability of open-source repositories. Some of the hyperlinks present in the CVE details page\footnote{\url{https://docs.qualcomm.com/product/publicresources/securitybulletin/february-2025-bulletin.html}} provided access to the vulnerable code and their patches. Moreover, some of the older code repositories were migrated from one host to another, which made the repository links inaccessible. The final list contained \textbf{180} instances of ASoCS vulnerabilities. 

\end{enumerate}



\subsection{Step 3: Vulnerability Mappings}
\label{subsec:step3}

 \subsubsection{Mapping - The root cause of the vulnerabilities}

A manual analysis process was followed  to map the vulnerabilities to their root causes. To make the process as objective as possible, each of the 180 vulnerabilities was assigned to two researchers (one from industry and another from academia). 
The researchers inspected the source code and commit messages independently for the assigned CVEs. While a CVE could theoretically stem from multiple contributing factors, for the purpose of this study, each vulnerability instance was mapped to a single, most probable root cause. This decision was made to maintain consistency and reduce ambiguity during manual classification.  
We initially followed the taxonomy of root causes described in \cite{ref_3} to analyze the ASoCS vulnerabilities. Later, we updated the taxonomy by using an open-coding technique to customize the list of root causes. Once each researcher mapped CVEs against their root causes, they had an internal discussion to eliminate the differences and decide on a single mapping version.

 \subsubsection{Mapping - Software Architectural Modules}

The ASoCS architecture model has already been described in \emph{Step 1}. We tried to map each vulnerability to the software modules present in the architecture model. A four-tier mapping scheme was followed for the purpose, i.e., each vulnerability was mapped to the main software module followed by three sub-modules in the hierarchy. For instance, \emph{CVE-2024-33049 was mapped as Central Processor (CP) $\rightarrow$ Kernel Space $\rightarrow$ WLAN Module $\rightarrow$ WLAN Roaming and Scan Management Module.} 
The process followed for the mapping, involved inspecting the location of the vulnerable file in the code repository and further examining the code content manually. This helped to understand its main functionality, thereby deciding the suitable sub-module that it should be a part of. Continuing with the same example, \emph{CVE-2024-33049}, we find that the associated file is \emph{umac/scan/dispatcher/src/wlan\_scan\_utils\_api.c} \footnote{\url{https://git.codelinaro.org/clo/la/platform/vendor/qcom-opensource/wlan/qca-wifi-host-cmn/-/commit/a469aec8aaae97c56b59a607308755348e0acc5f}}. 
In the following, we explain through guiding questions how we placed this example vulnerable file in a particular software module. 

\begin{itemize}
    \item What is the purpose of the file?

The \emph{wlan\_scan\_utils\_api.c} file controls Wi-Fi scanning, network discovery, and connection management. It also handles scanning for access points (APs) required during network discovery, handover, and smooth roaming services (802.11k, 802.11v, 802.11r \cite{ref_14}).

\item Why is the file, part of the CP and not a separate Wi-Fi module?  

Wi-Fi is part of the OS networking stack, not a separate module. Moreover, scanning needs direct interaction with the OS (e.g., Linux) networking services.

\item Why is the file part of the \emph{Kernel} space and not \emph{User} space?

User space components can request a scan but cannot directly access or control WLAN hardware. This is mainly handled by the Kernel WLAN driver, which executes actual Wi-Fi scans, manages connections, and interacts with the WLAN firmware.
\end{itemize}
This example shows that we track the file that contains the vulnerable code, look into its details manually, and use this information to  find the best associated location in the architecture model. 



Again, this mapping was done independently by two researchers, with the deviations discussed to resolve differences and create a common mapping.

 \subsubsection{Mapping - Vulnerability Mitigation Time}

Tracking vulnerability mitigation time is critical in ASoCS security analysis \cite{ref_141} because delays in patching can leave safety-critical systems exposed to exploitation, increasing risk over time. Understanding these delays helps prioritize hardening efforts and improve response strategies for high-impact components. Therefore in this study, we noted the vulnerability mitigation time for all the \textbf{180} vulnerable codes and later mapped the same across the different software modules and CWEs. Formula \ref{eq1} was used to calculate the mitigation time (MT) in days.

\begin{equation} \label{eq1}
MT =
\begin{cases}
GCD - RD, & \text{if } RD < GCD \\
1, & \text{if } GCD = GAD \text{ and } RD > GCD \\
GCD - GAD, & \text{if } RD > GCD
\end{cases}
\end{equation}

Here, GCD= Git Commit Date (The timestamp when the commit was merged or applied into the public repository (i.e., made publicly visible).), GAD= Git Commit Authored Date (The timestamp when the patch commit was authored (i.e., originally written by the developer).) and RD= Reported Date (The date the vulnerability was published in the NVD or disclosed via official security bulletin)


The mitigation time is usually calculated as the number of days between the GCD and RD. Under exceptional circumstances like the date of reporting is after the commit date, the vulnerability is assumed to be found internally, and in all such cases, the mitigation time is calculated as the number of days between GCD and GAD. In case GAD and GCD are on the same date, we consider the mitigation time to be 1 day (to avoid zero valued entries)

The affected chipsets and CWEs were accessed directly from hyperlinks provided on the NVD CVE details page\footnote{\url{https://nvd.nist.gov/vuln/detail/CVE-2024-38416}}. 
\section{Results and Analysis}
\label{sec:results}
Our final results include a list of \textbf{180} ASoCS vulnerabilities with all the mappings described in Section \ref{sec:methodology}. The full list of vulnerabilities can be accessed in our online repository \footnote{\url{https://github.com/SriAbir/AutomotiveSoCSoftware}}. In this section, we answer our research questions by presenting different relationships between the CVEs, CWEs, software modules, and the vulnerability mitigation time for our data. 

\textbf{RQ1}: What are the root causes behind the ASoCS vulnerabilities, and to what extent can these root causes explain the identified vulnerabilities?

Table \ref{tab1} presents the list of the top 10 out of the 16 root causes identified against 180 CVEs.

\begin{table}[htbp]
\caption{Root Causes of ASoCS vulnerabilities} 
\tiny
\begin{center}
\begin{tabular}{|p{0.30\linewidth} | p{0.60\linewidth} | }
\hline
\textbf{Root Cause}  & \textbf{Description}\\
\hline
Missing Size/Length Validation  & This includes all missing validations for a variable, array, pointer, etc., size (e.g., registers, buffers, etc.) \\
\hline
Improper Condition Checks  & This can include faulty conditional expressions in the program, which may or may not be a part of a looping statement \\
\hline

Improper Concurrency Controls  & This involves misuse of concurrency-related structures (e.g., mutex, critical regions, threads, etc.). \\
\hline
Improper Algorithm Implementation & This can cover a wider range of problems where the algorithm was not designed properly (missing or faulty decision logic) \\
\hline
Missing Checks for Memory Duplication/Unmapping  & This includes missing checks for proper release and usage of memory buffers. Generally responsible for out-of-bounds and overflow problems.  \\
\hline
Missing Access Control Mechanism  & This can include missing privilege or permission (read, write, execute) for a data structure \\
\hline
Improper Sequence of Operations  & The proper sequencing of codes can be a major issue in some cases (e.g., incrementing a counter before and not after an operation) \\
\hline
Improper Variable Scoping  & This includes problems related to the usage of global and process-specific, i.e., local data structures \\
\hline
Missing Assignment  & This can include missing initialization or re-initialization of variables, improper array size assignment, etc. \\
\hline
Missing Session Validation  & This includes missing checks for device and link handles, also to verify the session they belong to. \\






\hline
\end{tabular}
\label{tab1}
\end{center}
\end{table}
The findings show that \textbf{Missing Size/Length
Validation} was the most common root cause, accounting for \textbf{72} out of 180 CVEs, followed by \textbf{Improper Condition Checks (25)} and \textbf{Improper Concurrency Controls (20)}.


In Fig.~\ref{r2}, we analyzed the effect of each of these root causes across the identified software modules using a sankey diagram. The main observations from Fig. \ref{r2} are as follows:



\begin{figure}[ht]

\includegraphics[scale=.45]{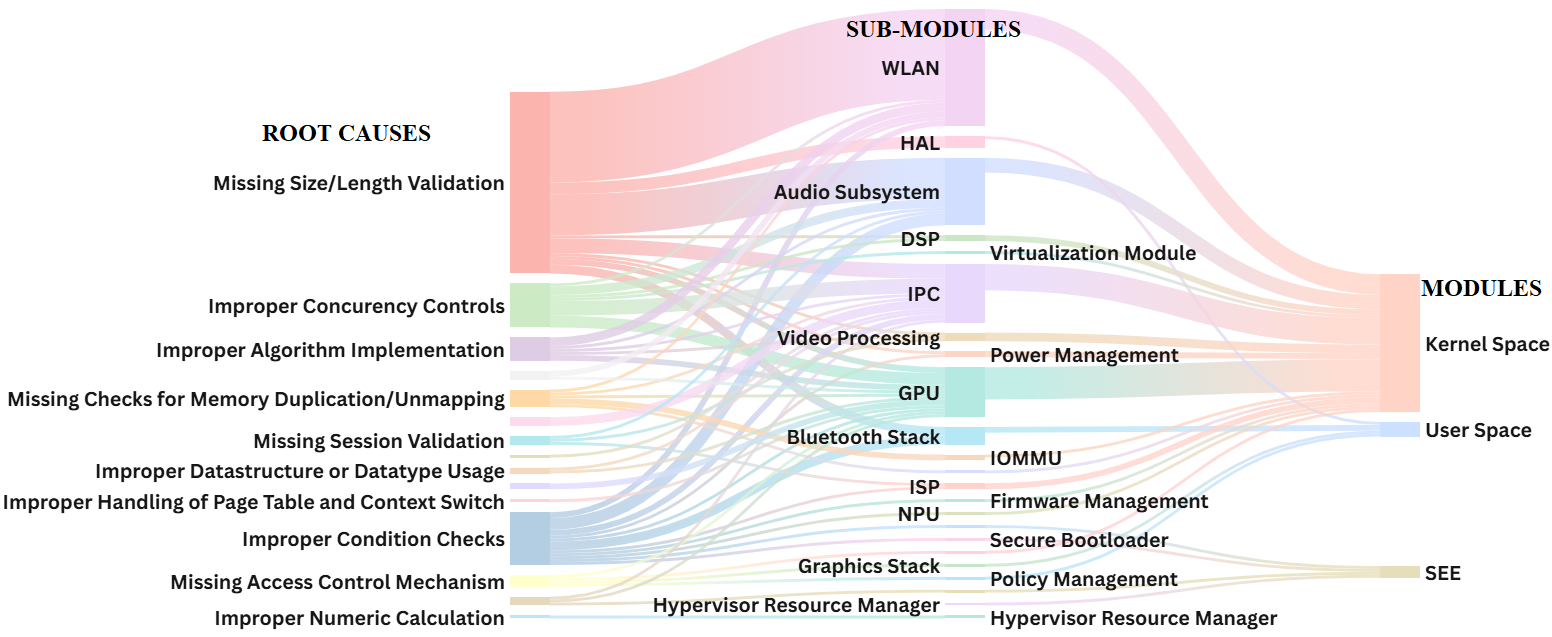}
\caption{Distribution of Root Causes across Software Modules}
\label{r2}

\end{figure}




\begin{itemize}

   \item \textbf{31} and \textbf{14} CVEs are caused by \textit{Missing Size/Length
Validation} in the \textit{WLAN} and \textit{Audio Subsystem} modules respectively (represented by thick flow lines) 

 \item The \textit{GPU Module} shows the maximum variety of associated root causes, \textbf{11} followed by \textit{IPC (Inter-process Communication), and WLAN modules} associated with \textbf{9} and \textbf{7} root causes respectively (indicated by multiple flow lines).

 \item The root cause \textit{Improper Condition Checks} is spread across maximum number of modules (\textbf{11}) followed by \textbf{Missing Size/Length Validation} and \textit{Improper Concurrency Controls} spread across \textbf{9} and \textbf{6} software modules, respectively
\end{itemize}

We also mapped how these root causes were responsible for the CWE categories for the identified ASoCS vulnerablities.  Fig.~\ref{r3} presents a network graph depicting the relationship between the CWEs (purple nodes) and the root causes (multi-color nodes). The main observations from Fig.~\ref{r3} are as follows:

\begin{figure}[ht]
\begin{center}
\includegraphics[scale=.40]{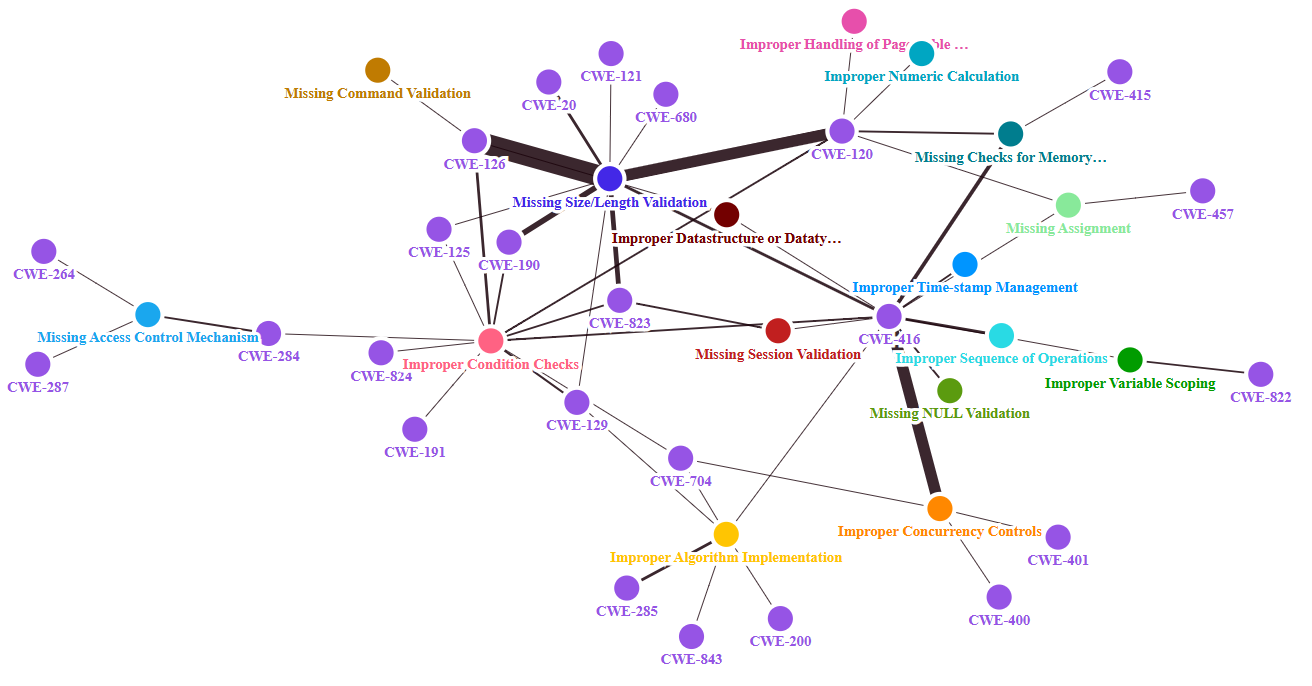}
\caption{Association of Root Causes with CWEs}
\label{r3}
\end{center}
\end{figure}

\vspace{-0.5cm}
\begin{itemize}

   \item There are \textbf{22} and \textbf{12} instances of \textit{CWE-129, Improper Validation of Array Index} and \textit{CWE-120, Buffer Copy without Checking Size of Input ('Classic Buffer Overflow')} respectively caused by \textit{Missing Size/Length Validation} root cause (shown by thick edges).

 \item There are \textbf{11} and \textbf{10} different CWEs associated with \textit{Improper Condition Checks} and \textit{Missing Size/Length Validation} respectively (depicted by the number of edges associated with the nodes). 

 \item \textit{CWE-416, Use After Free} can happen due to a wide variety of \textbf{12} root causes followed by \textit{CWE-120, Buffer Copy without Checking Size of Input ('Classic Buffer Overflow')} with \textbf{8} root causes.

\end{itemize}

The findings from \textbf{RQ1} reveal the frequently occurring root cause in ASoCS and their association with different software modules and CWE categories. This dual-layered mapping not only reveals which architectural components are most prone to specific types of flaws, but also helps contextualize these issues within standardized vulnerability classifications. Such insight is crucial for guiding module-specific hardening efforts, informing secure development practices, and aligning remediation strategies with known weakness patterns across the ASoCS stack.



\textbf{RQ2}: How are the vulnerabilities distributed across different architectural components/modules of the ASoCS architecture model stack?

Our findings show that the kernel space, user space, and secure execution environment account for \textbf{86.6\%}, \textbf{11.9\%} and \textbf{1.5\%} of the total ASoCS vulnerabilities, respectively.  Fig.~\ref{r4} further elaborates on the distribution of CVEs across the kernel space. As evident from the tree map, \textit{WLAN \textbf{(42)}, Audio Subsystem \textbf{(22)}, IPC (Inter-Processor Communication) \textbf{(19)} and GPU \textbf{(18)}} modules contribute most towards the total CVE count. Each of these modules are further expanded and it is seen that the \textit{WLAN Roaming \& Scan Management \textbf{(16)}, MLO Manager \textbf{(12)}} and \textit{ALSA \textbf{(12)}} sub-modules have the highest CVE counts.

\begin{figure}[ht]
\begin{center}
\includegraphics[scale=.50]{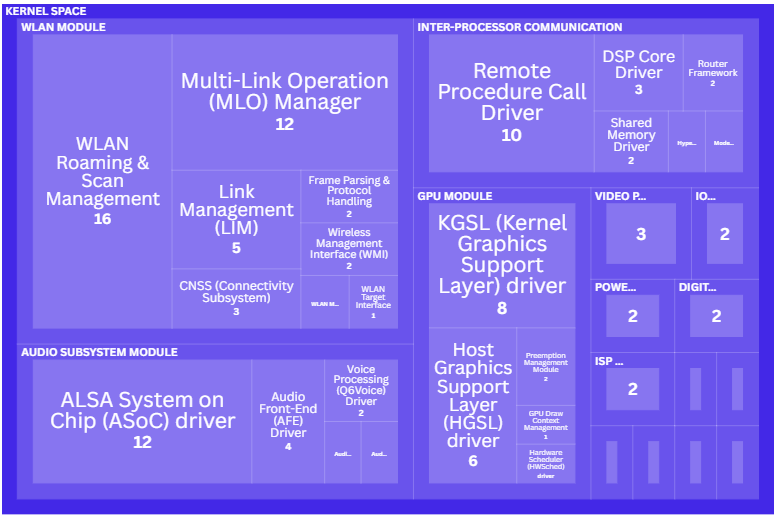}
\caption{Distribution of CVEs across Software Modules}
\label{r4}
\end{center}
\end{figure}


Fig.~\ref{r5} presents the CWE distribution across the most populated modules with the help of a heat map. Here, a red cell symbolizes the maximum occurrence of a particular CWE type in the module. As the color goes from light to darker blue, the number of CWEs gradually decreases. The main observations from the heatmap are as follows:

\begin{itemize}

   \item  The spread of \textbf{9} CWEs across the \textit{GPU module}, followed by \textbf{8} CWEs in \textit{WLAN} and \textit{Audio Subsystem} each and \textbf{7} in \textbf{IPC}.

 \item The \textit{WLAN module} is dominated by \textit{CWE-126, Buffer Over-read \textbf{(25 instances)}}, \textit{Audio subsystem} module with \textbf{7} instances of \textit{CWE-120, Buffer Copy without Checking Size of Input ('Classic Buffer Overflow')} and \textit{GPU and IPC module} with \textbf{9} and \textbf{8}  cases of \textit{CWE-416, Use After Free} respectively.

\end{itemize}

The answers for \textbf{RQ2} provide the frequency and distribution of the CVEs and CWEs across the ASoCS architectural stack.  These findings can help system architects and developers to prioritize security reviews, refactoring, or isolation strategies in these critical areas. Furthermore, such mapping supports a layered defense approach, enabling targeted mitigation efforts aligned with the structural roles of components within the SoC stack.

\begin{figure}[ht]
\begin{center}
\includegraphics[scale=.25]{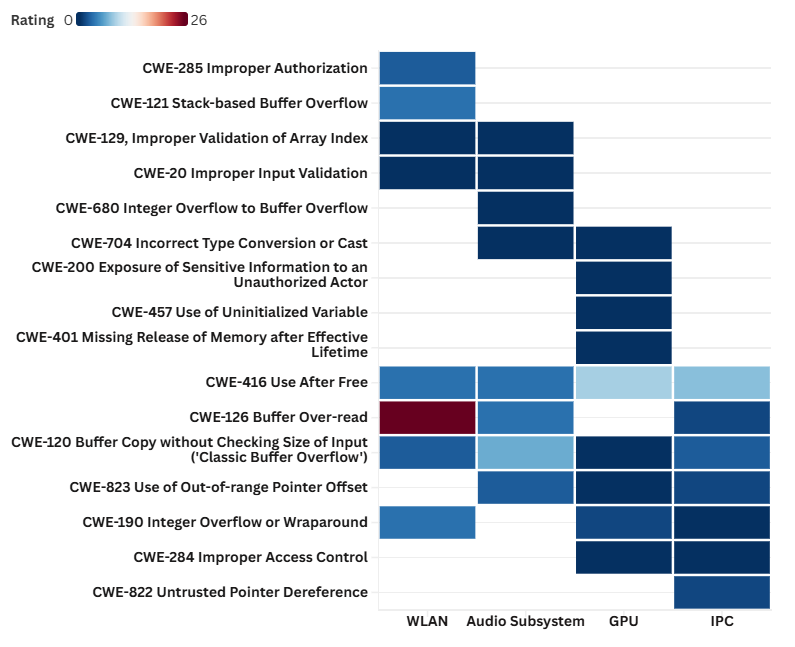}
\caption{Distribution of CWEs across Software Modules}
\label{r5}
\end{center}
\end{figure}
\textbf{RQ3}: How does the vulnerability mitigation time vary across different Common Weakness Enumeration (CWEs) and ASoCS modules?

Results show that \textit{CWE-416, Use After Free} and \textit{CWE-126, Buffer Over-read} contribute towards \textbf{25.4\%} and \textbf{24.5\%} of the total vulnerability mitigation time. On the other hand,  \textbf{26.4\%} and \textbf{23.2\%} of the total mitigation time were found to be spent on \textit{WLAN and IPC modules} respectively.

We further analyzed the variation of the mitigation time (in days) across some of the most frequently occurring CWEs. Fig.~\ref{r7}.a) presents the same in the form of violin plots. \textit{CWE-120} and \textit{CWE-190} show a lower variation of mitigation time (here, all the CVEs are mitigated within 1 to 3 months). In the case of \textit{CWE-416}, though most of the CVEs are mitigated within 1 - 3 months, a few take a longer time of 5 - 7 months. For \textit{CWE-126}, most CVEs get resolved within 1-4 months, except 1 taking 11 months. Finally, \textit{CWE-823} shows a wider spread along the curve, indicating different mitigation times for different instances of the CWE (the lowest being 1 month and the highest almost 10 months).

\begin{figure}[ht]
\begin{center}
\includegraphics[scale=.55]{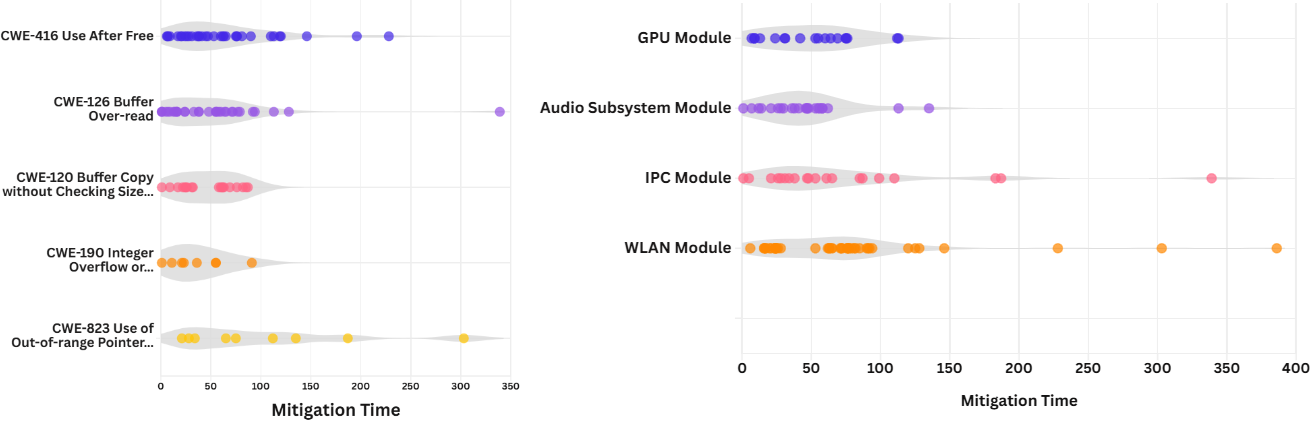}
\caption{Variation of Vulnerability Mitigation Time across top) CWE \& bottom) Modules}
\label{r7}
\end{center}
\end{figure}
We also explored how the mitigation time varies across some of the most vulnerable software modules. Fig.~\ref{r7}.b) presents the violin plot, showing that the \textit{Audio Subsystem} module has the lowest variation, with the maximum number of CVEs being mitigated within 1 -- 2 months. The next lowest variation is the \textit{GPU} module with a slight variation (most of the CVEs get mitigated within 3.5 months, with a few extending to 4.5 months). A wider variation is seen in the \textit{WLAN and IPC} modules where mitigation time ranges from 
(1 day to approximately 1 year). 


When addressing \textbf{RQ3} we found that mitigation time analysis across CWEs and architectural modules may provide valuable insight into the operational security health of an ASoCS stack, especially important in safety-critical CPS systems like vehicles. Modules exhibiting shorter patch delays often reflect well-maintained components with active developer engagement, efficient CI/CD integration, or simpler dependency structures. In contrast, consistently longer mitigation times may indicate areas of architectural complexity, insufficient ownership, or low visibility within the maintenance workflow. Mapping these delays helps identify risk-prone modules where delayed patching could leave safety-critical systems vulnerable for extended periods. These insights can guide security hardening priorities, promote architectural refactoring, and contribute to more resilient software lifecycle management by addressing architectural areas with slower security response in CCAM.

\begin{boxA}
\textbf{Summary}: Missing Size/Length Validation was found to be the most dominant root cause for the ASoCS. The WLAN module contributed to 37\% of the total vulnerabilities (main category: CWE-126). The GPU module showed the maximum variety of CWE categories. CWE-416 consumed the maximum time for mitigation. WLAN and IPC modules showed a wide variation in mitigation time.
\end{boxA}
\section{Discussion}
When  visualized from the perspective of CPS, these findings raise critical concerns. Considering that the automotive industry is quickly moving towards fully centralized SoCs with software-defined vehicle architecture, the CVE of a single software module can be easily exploited by an attacker to gain access of the entire system in the worst case. For example, consider a hacker controlled Road Side Unit (RSU) or an Access Point sending a malicious beacon frame (MBF) \cite{ref_15} to the SoC Wi-Fi Driver (part of \emph{WLAN module} of our model). Due to an existing vulnerability (of type CWE-126, Buffer Over-read) in the \emph{wlan\_mlo\_t2lm.c} file, the module would accept the MBF without any validation. This could lead to various security issues like data leakage due to uninitialized memory read, crashing of the Wi-Fi stack leading to DDoS (Distributed Denial of Service), traffic signal communication (V2I) failure, collision avoidance alerts (V2V) failure, fleet management and over-the-air (OTA) updates failure, etc.


Although our architecture was not constructed directly from AUTOSAR blueprints, the structural and service-layer similarity with AUTOSAR Adaptive Platform allows us to extrapolate potential integration and migration use cases. E.g, for organizations adopting AUTOSAR, particularly the adaptive platform, our findings offer a grounded perspective on the real-world vulnerabilities encountered in SoC-based systems. Imagine a scenario where a Tier-1 automotive supplier is transitioning its infotainment and connectivity Electronic Control Unit (ECU) from a proprietary RTOS to an AUTOSAR Adaptive Platform. The ECU integrates a Snapdragon-based SoC and supports vehicle-to-infrastructure (V2I) features via Wi-Fi. In such a situation our findings could help the security lead to assess the modules that pose the highest security risks and where mitigation or architectural hardening should be prioritized. Depending on the assessment, the company can take different integration hardening decisions, targeted fuzz testing and buffer boundary checks, during third-party software procurement, insists on memory-safe implementations for the Wi-Fi driver, etc.

\section{Threats to Validity}
\label{sec:validity}

 We used the frameworks by Wohlin \cite{ref_18} and Staron \cite{ref_19} while considering the validity of our study.

In the \emph{construct validity} category, the main risk was missing important ASoCS vulnerabilities. Both, our study methodology and restriction to open-source SoC software could have resulted into missing vulnerability instances in some of the most popular components of ASoCS like ADAS or Digital Cockpit \cite{ref_17}.  Construct validity may also be affected by the inclusion/exclusion of real-time vs high-end modules.

In the \emph{conclusion validity} area, we see our main risk as our manual mapping process. Although we considered using automated tools, e.g., LLMs, we chose to do it manually to be able to validate our results or discuss them if needed. 

The study of mitigation timelines may relate to a module’s architectural role or integration depth, but we also acknowledge that open-source project dynamics, such as contributor activity and governance can influence these outcomes, which represents a potential threat to internal validity.

A key limitation as \emph{external validity} is that the ASoCS architecture model was manually derived from open-source repositories, which may not represent all variations of commercial SoC implementations. While we aligned our model with AUTOSAR concepts, we acknowledge that there is a lack of formal validation (e.g., expert interviews or industry-confirmed mappings). 



\section{Conclusion and Future Work}

\label{sec:conclusion}

Automotive SoC Software architecture is becoming more centralized and complex with time. While this evolution supports benefits like efficient OTA updates as well as further customer services provided via cloud environments, seamless data sharing, and optimized power consumption, it also amplifies the security risks inherent to tightly integrated CPS platforms. We provide in this study an insight of some of the most vulnerable regions of the ASoCS architecture stack. By mapping 180 real-world vulnerabilities to software modules, CWE categories, root causes, chipsets, and mitigation delays, we identified critical trends such as vulnerability concentration in the Wi-Fi module and frequent occurrences of missing size/length validation flaws.



Derived from publicly available repositories and aligned with AUTOSAR principles, our model helps identify vulnerable components that may be overlooked in conventional security audits. Security testers and architects adopting AUTOSAR can leverage our root cause mappings and delay patterns to prioritize high-risk modules, guide targeted hardening, and validate integration decisions during platform adoption. Future work includes validating the architecture model with AUTOSAR practitioners and domain experts as well as expanding attack path analysis to support broader CPS domains.

\scriptsize {The data used in this study can be accesses at
\url{https://github.com/SriAbir/AutomotiveSoCSoftware}.}

%

\begin{credits}
\subsubsection{\ackname} This paper has been partially financed by Software Center, www.software-center.se and SFO Transport/AoA Transport at the University of Gothenburg and Chalmers University of Technology. 

\end{credits}
%
%
%

\begin{thebibliography}{8}



\balance
\bibitem{ref_1}
Dark-Reading,\url{https://www.darkreading.com/cyberattacks-data-breaches/volkswagen-breach-exposes-data-of-800k-customers}, last accessed 2025/03/17


\bibitem{ref_3}
Garcia, J., Feng, Y., Shen, J., Almanee, S., Xia, Y. and Chen, A.Q.A.: A comprehensive study of autonomous vehicle bugs. In Proceedings of the ACM/IEEE 42nd international conference on software engineering pp. 385-396. (2020)

\bibitem{ref_4}
Mashkoor, A., Egyed, A., Wille, R. and Stock, S. :Model‐driven engineering of safety and security software systems: A systematic mapping study and future research directions. Journal of Software: Evolution and Process, 35(7), p.e2457. (2023)
\bibitem{ref_5}
Bella, G., Biondi, P., Costantino, G. and Matteucci, I., 2022. Designing and implementing an AUTOSAR-based Basic Software Module for enhanced security. Computer Networks, 218, p.109377.

\bibitem{ref_20} 
Basu, S, Staron, M. :Understanding the Changing Landscape of Automotive Software Vulnerabilities: Insights from a Seven-Year Analysis. Accepted in SVM Workshop (ICSE 2025), \url{https://arxiv.org/abs/2503.17537} (2025)

\bibitem{ref_6}
Codelinaro, qcom2, \url{https://git.codelinaro.org/clo/la/kernel/qcom2}, last accessed 2025/03/17

\bibitem{ref_7}
Qualcomm Linux Landing Page, \url{https://docs.qualcomm.com/bundle/publicresource/topics/80-70014-115/overview.html}, last accessed 2025/03/17

\bibitem{ref_8}
AUTOSAR, Explanation of Adaptive Platform Software Architecture, 2022. May. 2025. [Online]. Available: \url{https://www.autosar.org/fileadmin/standards/R22-11/AP/AUTOSAR\_EXP\_SWArchitecture.pdf},

\bibitem{ref_9}
AUTOSAR, The Standardized Software Architecture. May. 2025. [Online]. Available: \url{https://gi.de/informatiklexikon/autosar-the-standardized-software-architecture}, last accessed 2025/03/17

\bibitem{ref_92}
AUTOSAR, Explanation of Security Overview, 2022 . May. 2025. [Online]. Available: \url{https://www.autosar.org/fileadmin/standards/R22-11/FO/AUTOSAR\_EXP\_SecurityOverview.pdf}






\bibitem{ref_14}
Sanchez, M.I. and Boukerche, A. :On IEEE 802.11 K/R/V amendments: Do they have a real impact?. IEEE Wireless Communications, 23(1), pp.48-55. (2016)

\bibitem{ref_141}
Ruohonen, J., 2024. An empirical study of vulnerability handling times in CPython. arXiv preprint arXiv:2411.00447.

\bibitem{ref_15}
Vanhoef, M., Adhikari, P. and Pöpper, C. :Protecting wi-fi beacons from outsider forgeries. In Proceedings of the 13th ACM Conference on Security and Privacy in Wireless and Mobile Networks pp. 155-160. (2020)

\bibitem{ref_18} 
Wohlin, C., Runeson, P., Höst, M., Ohlsson, M.C., Regnell, B. and Wesslén, A. :Experimentation in software engineering (Vol. 236). Berlin: Springer. (2012)

\bibitem{ref_19} 
Staron, M. :Action research in software engineering (p. 50). Berlin: Springer International Publishing. (2020)

\bibitem{ref_17}
Qualcomm Snapdragon TM Automotive Platform S820A, \url{https://www.qualcomm.com/content/dam/qcomm-martech/dm-assets/documents/820a-product-sheet-v3\_0.pdf}, last accessed 2025/03/18


\end{thebibliography}
%

\end{document}